\documentclass[11pt]{article}
\usepackage[utf8]{inputenc}
\usepackage[margin=1in]{geometry}
\usepackage{graphicx,amssymb}
\usepackage{microtype}
\usepackage{ulem}
\usepackage{caption}
\usepackage{subcaption}
\usepackage{multicol}
\usepackage{hyperref}
\usepackage{soul}
\usepackage{multirow}
\usepackage{xcolor, amsmath, textcomp}
\usepackage{authblk}
\usepackage{lineno}
\usepackage{array}
\usepackage{makecell}
\setcellgapes{2pt}

\newcommand{\vg}{v_g}
\newcommand{\Px}{P(\Delta x)}
\newcommand{\Fi}{\vec{F}_{SA}}

\title{Noisy circumnutations facilitate self-organized shade avoidance in sunflowers}

\author[1]{Chantal Nguyen}
\author[2]{Imri Dromi}
\author[2]{Aharon Kempinski}
\author[2-4]{Gabriella E. C. Gall}
\author[1,5-9]{Orit Peleg$^{\dag}$}
\author[2]{Yasmine Meroz$^\ddag$}
\affil[1]{BioFrontiers Institute, University of Colorado, Boulder, CO, USA}
\affil[2]{School of Plant Science and Food Security, Tel Aviv University, Tel Aviv, Israel}
\affil[3]{Zukunftskolleg, Department of Biology, University of Konstanz, Konstanz, Germany}
\affil[4]{Centre for the Advanced Study of Collective Behaviour, Konstanz, Germany}
\affil[5]{Department of Computer Science, University of Colorado, Boulder, CO, USA}
\affil[6]{Department of Physics, University of Colorado, Boulder, CO, USA}
\affil[7]{Department of Applied Math, University of Colorado, Boulder, CO, USA}
\affil[8]{Department of Ecology and Evolutionary Biology, University of Colorado, Boulder, CO, USA}
\affil[9]{Santa Fe Institute, Santa Fe, NM, USA}
\begin{document}
\maketitle

\begin{abstract}
Circumnutations are widespread in plants and typically associated with exploratory movements, however a quantitative understanding of their role remains elusive. In this study we report, for the first time, the role of noisy circumnutations in facilitating an optimal growth pattern within a crowded group of mutually shading plants. We revisit the problem of self-organization observed for  sunflowers, mediated by shade response interactions. Our analysis reveals that circumnutation movements conform to a bounded random walk characterized by a remarkably broad distribution of velocities, covering three orders of magnitude. In motile animal systems such wide distributions of movement velocities are frequently identified with enhancement of behavioral processes, suggesting that circumnutations may serve as a source of functional noise. To test our hypothesis, we developed a parsimonious model of interacting growing disks, informed by experiments, successfully capturing the characteristic dynamics of individual and multiple interacting plants. Employing our simulation framework we examine the role of circumnutations in the system, and find that the observed breadth of the velocity distribution confers advantageous effects by facilitating exploration of potential configurations, leading to an optimized arrangement with minimal shading. These findings represent the first report of functional noise in plant movements, and establishes a theoretical foundation for investigating how plants navigate their environment by employing computational processes such as task-oriented processes, optimization, and active sensing. (218 words)

\end{abstract}

% \begin{keyword}
plant tropism; circumnutation; shade avoidance;  collective behavior; self-organization; sunflowers; minimal model
% \end{keyword}

% \begin{teaser}
    \textbf{One sentence summary of paper.}The study highlights noisy circumnutations as a strategy plants use for optimizing growth in crowded conditions.
% \end{teaser}

% \begin{corrauthor}
\noindent $^\dag$orit.peleg@colorado.edu \\
\noindent $^\ddag$jazz@tauex.tau.ac.il \\
% \end{corrauthor}

\section*{Introduction}

The survival of plants greatly depends on light availability. In many natural habitats multiple neighboring plants shade each other, competing over this critical resource. The presence of neighbors varies over space and time, and plants have evolved the ability to detect neighbors and respond by adapting their morphology accordingly~\cite{Callaway2003, Schmitt2003, Schmitt1993, Pigliucci2001, Gruntman2017}. 
Indeed, a fundamental difference between plants  and other motile organisms is that plants generally move by growing; an irreversible process which imbues plant movement with a commitment to the permanent morphology. 
The direction of plant growth is dictated either by external directional cues such as light,  processes termed tropisms, or by  inherent internal cues, such as the exploratory periodic movements termed \textit{circumnutations}. 

Recently Pereira et al.~\cite{Pereira2017} found that sunflower crops growing in a row at high densities self-organized into a zig-zag conformation of alternating inclined stems, thus collectively increasing light exposure and seed production. Self-organized processes refer to initially disordered systems where order arises from local interactions between individuals, facilitated by random perturbations, or noise.
Local interactions were found to be mediated by the shade avoidance response, a form of tropism where plant organs grow away from neighboring plants~\cite{Casal2013}, responding to changes in the ratio between red and far-red wavelengths, characteristic of the  light spectrum of plant shade
~\cite{Ballare1992, Maddonni2002, Novoplansky1990}. However, the source of perturbations required for the observed self-organization in this system -- remains elusive. 

Noise plays a critical role in self-organized systems: 
at the right magnitude relative to the interactions, it enables the system to explore a variety of states, thus enabling reliable adaptation to short-term changes in the environment while maintaining a generally stable behavior~\cite{Helbing2002,Helbing1999,vicsek1999, Kahneman2022}. Too little noise confines the system to a sub-optimal state, while too much noise masks the interactions.
In biological systems, noise is often identified as being \textit{functional}, exhibiting 
a wide spectrum of manifestations.
For example, organisms use noise in order to increase sensory salience, ultimately enabling them to balance the behavioral conflict between producing costly movements for gathering information (“explore”) versus using previously acquired information to achieve a goal (“exploit”)~\cite{Padilla2014,Wilson2021,Biswas2023}. 
Additionally, the navigation paths of bacteria \cite{wadhwa_bacterial_2022}, insects \cite{peleg_optimal_2016}, and mammals \cite{fortin_elk_2005,flossmann_spatial_2021} exemplify the intricate trajectories adopted to counterbalance uncertain surroundings \cite{pfuhl_precision_2010}. 
Likewise, the behavioral variability observed in honeybees, specifically in the triggering of fanning, serves as a mechanism for ventilating their hives \cite{peters_collective_2019}. In all of these instances, noisy processes play a fundamental role in facilitating biological functionality.

Here we propose cirumnutations as the predominant source of perturbations in this system.
These movements are ubiquitous in plants, following elliptical or irregular trajectories with large variations in both amplitude and periodicity~\cite{darwin_2009,hart_plant_1990,kiss_up_2006,Mugnai2007, baillaud1962}.
Circumnutations assist climbing plants in locating mechanical supports~\cite{darwin_2009, larson2000circumnutation}, facilitate root navigation around obstacles~\cite{taylor_mechanism_2021,tedone_optimal_2020}, and contribute to the regulation of shoot stability during elongation growth and tropic bending~\cite{schuster_circumnutations_1997, bastien_coupled_2018}. However, their ecological function in non-climbing shoots, which constitute the majority of plants, remains unclear~\cite{migliaccio_circumnutation_2013,stolarz_circumnutation_2009,baillaud1962,larson2000}.

To test our hypothesis that noisy circumnutation can benefit collective plant growth, we develop a minimal data-driven model where mutually shading sunflower crowns are represented by interacting growing disks, and circumnutations  are represented by random perturbations with equivalent statistics. The model, informed and validated by our experimental data, recovers the observed dynamics of sunflower growth patterns, and enables us to examine the functionality of circumnutations. 
Our model finds that the characteristic statistics of measured circumnutations are such that they maximize the light exposure of the system,  
 suggesting that circumnutations play a critical role in reaching collectively optimal growth configurations under light competition. 

\begin{figure}[t!]
    \centering
    \includegraphics[width=\linewidth]{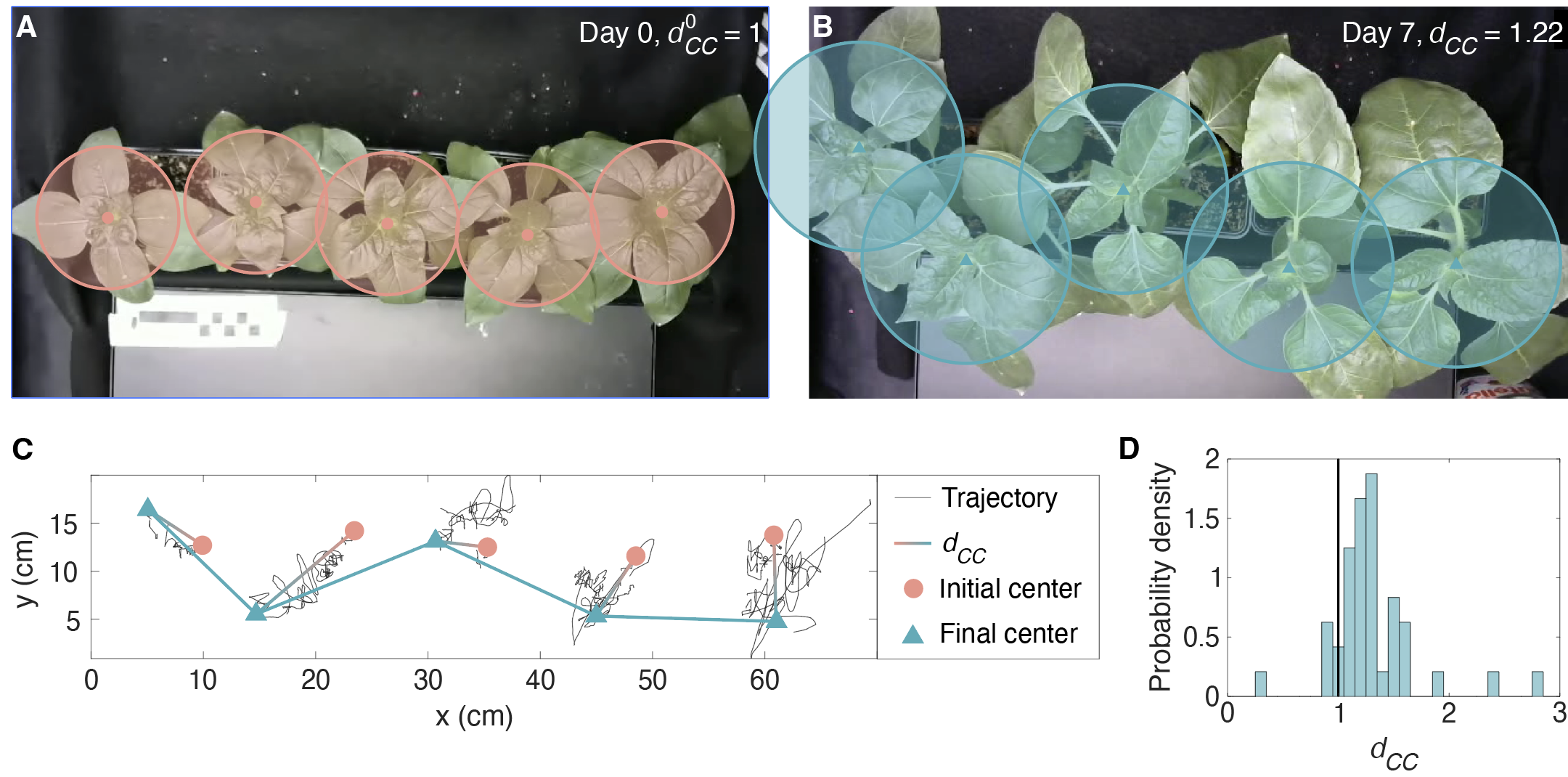}
    \caption{\textbf{Plants grown in a dense row deflect away from each other to minimize their mutual shading. (A)} Snapshot of sunflowers (\textit{Helianthus annuus})  placed in a row, on the first day of recording. Pink dots indicate crown centers, and circles illustrate the model representation of crowns as disks in the $x$-$y$ plane. The initial average center-to-center separation between pairs of adjacent plants is normalized such that $d^0_{CC}=1$ (Eq.~\ref{eq:c2c}). 
    \textbf{(B)} After seven days, plants deflect from the center line, captured by the increased center-to-center distance $d_{CC}=1.22$. Blue triangles indicate crown centers, and circles represent crowns. 
    \textbf{(C)} Trajectories of crown centers during a sample 7-day recording are shown (black lines), where initial and final crown positions are represented by pink dots and blue triangles respectively. Blue lines illustrate the increased center-to-center separation ${d}_{CC}$ between pairs of adjacent plants, highlighting the arising deflected pattern. 
    \textbf{(D)} Histogram of the final $d_{CC}$ values from 12 multiple-plant experiments, indicating a final separation greater than that of the initial separation.}
    \label{fig:plants_row}
\end{figure}

\begin{figure}[t!]
    \centering
    \includegraphics[width=\linewidth]{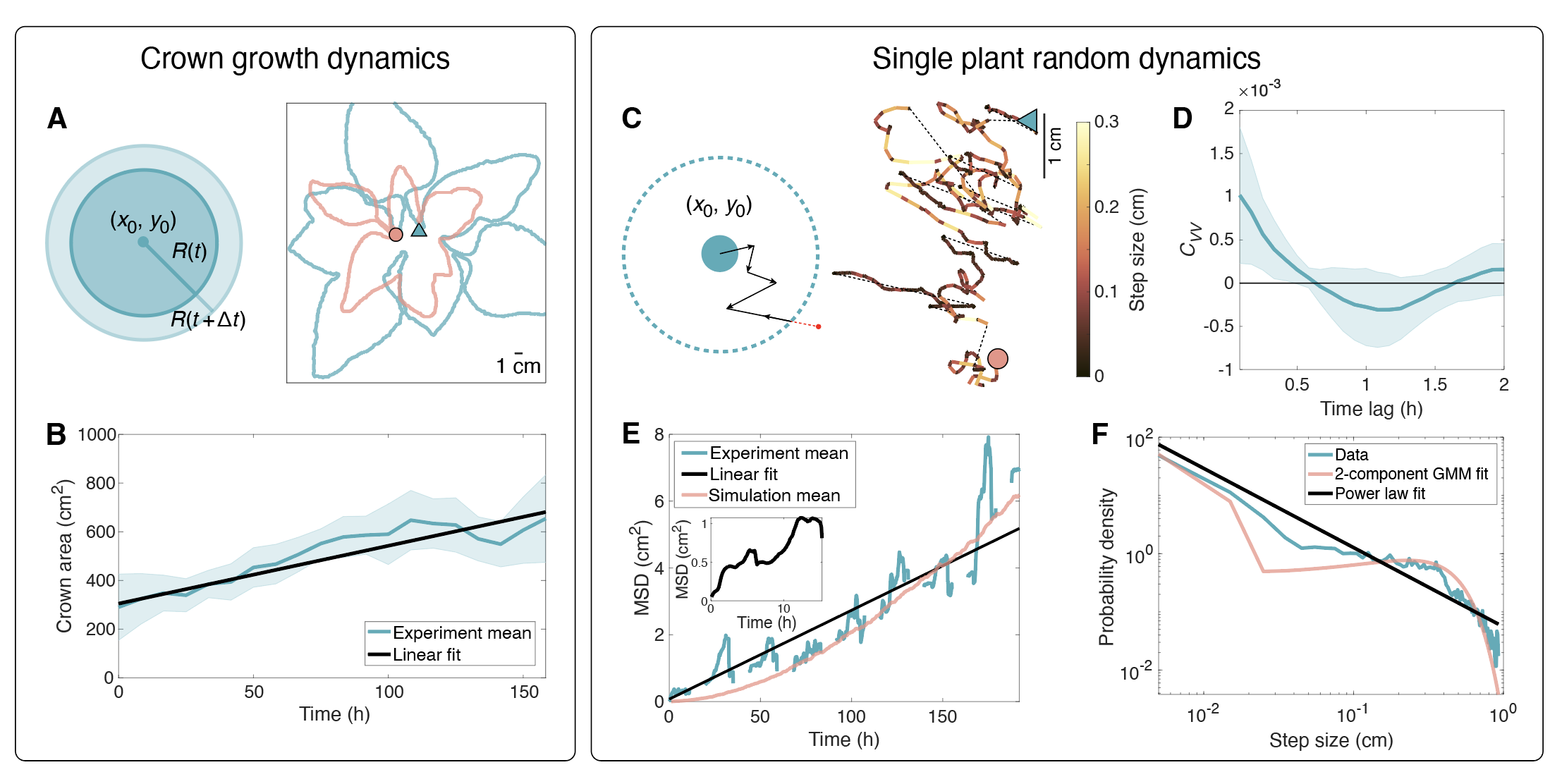}
    \caption{ \textbf{Characterization of single-plant dynamics. (Left) Crown growth dynamics (A)} Left: graphical representation of model; plant crown is approximated with a circle of radius $R(t)$ that increases according to the growth rate $v_g$ extracted from experiments (Eq.~\ref{eq:growth}, panel C). Right: example outlines of a segmented plant crown at the beginning (pink) and after 10 days (blue). \textbf{(B)} Crown area, $A(t)$, as a function of time, averaged over $N=8$ experiments; shaded area represents 1 standard deviation. Each experiment is individually fit to a line, the average fit (black) yields a slope $v_g = 2.4 \pm 0.5 \text{cm}^2/h$ (Eq.~\ref{eq:growth}). \textbf{(Right) Circumnutation dynamics} \textbf{(C)} Left: graphical representation of model; the crown (blue circle) is bounded by a reflective boundary (dashed line), increasing with time following Eq.~\ref{eq:boundary}. The crown follows a random walk (RW) and takes a step in a random direction at each time step; if a step crosses the boundary (red arrow) it is reflected back. 
   Right: example of measured crown trajectory. The initial and final positions of the crown, after 10 days, are marked by a pink dot and blue triangle accordingly. Gaps in the trajectory indicate the nightly 8h ``dark'' period (not recorded). Color code represents local velocity (light colors represent a larger step size, and therefore higher velocity). 
    \textbf{(D)} The autocorrelation of steps $C_{VV}$ as a function of time lag shows that at a time lag of $\Delta t = 0.6$ h, steps are uncorrelated.
    \textbf{(E)}
    The mean squared displacement (MSD) across the length of single plant experiments, averaged over all plants. The gaps in the curve represent 8-hour nighttime periods where the plant is not recorded. The black line represents the linear fit in Eq.~\ref{eq:MSD} with $R^2=0.84$. The pink line represents the mean MSD over 500 simulation instances. Each simulation follows the dynamics of a single plant in uniform light conditions. The daily experimental MSD averaged over single days (inset) exhibits a slight decrease of the slope after about 1 h, reflecting the limit on crown movement due to the length and flexibility of their stems.
    \textbf{(F)} The step size distribution (blue line). The sampling rate of trajectories is based on the autocorrelation of steps $C_{VV}$ (panel (D)). The resulting distribution of step sizes (equivalent to velocities) is wide, spanning three orders of magnitude, and can be approximated to a power law (black line) shown in Eq.~\ref{eq:step_dist}, or a 2-parameter Gaussian mixture model (GMM, pink line). While the GMM fit is slightly better, we choose the power law form since it allows us to describe the step behavior with a single parameter.}
    \label{fig:single_plant}
\end{figure}

\section*{Results}
\subsection*{Recapitulating alternate inclination experiments}

We recapitulate the self-organized zigzag growth conformation observed by Pereira \textit{et al.}~\cite{Pereira2017} in a controlled environment, enabling us to record the growth dynamics throughout. Fig.~\ref{fig:plants_row}A shows an initial arrangement of five young sunflowers (\textit{Helianthus annuus}) approximately 7 days old, where the plant crowns and their centers are highlighted, clearly aligned to a horizontal line. 
We define the position of crown $i$ at time $t$ as $\vec{r}_i(t)$, 
and define the average center-to-center distance between adjacent plants $d_{\text{CC}}(t)$ (shown in Eq.~\ref{eq:c2c}), normalized by the initial center-center distance at $t=0$  
so that $d_{\text{CC}}(0) = 1$ by definition. This value serves as a measure of how closely ($d_{\text{CC}}< 1$) or sparsely ($d_{\text{CC}}> 1$) the plant crowns are distributed. 
We allow the plants to grow undisturbed over 7 days, following the position and size of the crowns throughout. 
Fig.~\ref{fig:plants_row}B shows their final configuration, where the crowns have grown in size, and their centers are clearly in a staggered formation. In this particular example, the average distance has increased such that  
$d_{\text{CC}} = 1.22$, ascribed to the zigzag configuration. 
The trajectories of the tracked crown centers over the course of this specific multiple-plant experiment are shown in Fig.~\ref{fig:plants_row}C. 
Calculating over 12 such experiments, we observe that the final center-to-center distance is generally greater than the initial distance $d_{CC} > 1$ (Fig.~\ref{fig:plants_row}D), indicating that crowns indeed deflect away from one another in dense growth setups. 

\subsection*{Minimal model informed by experiments}
\begin{figure}[t!]
    \centering
    \includegraphics[width=\linewidth]{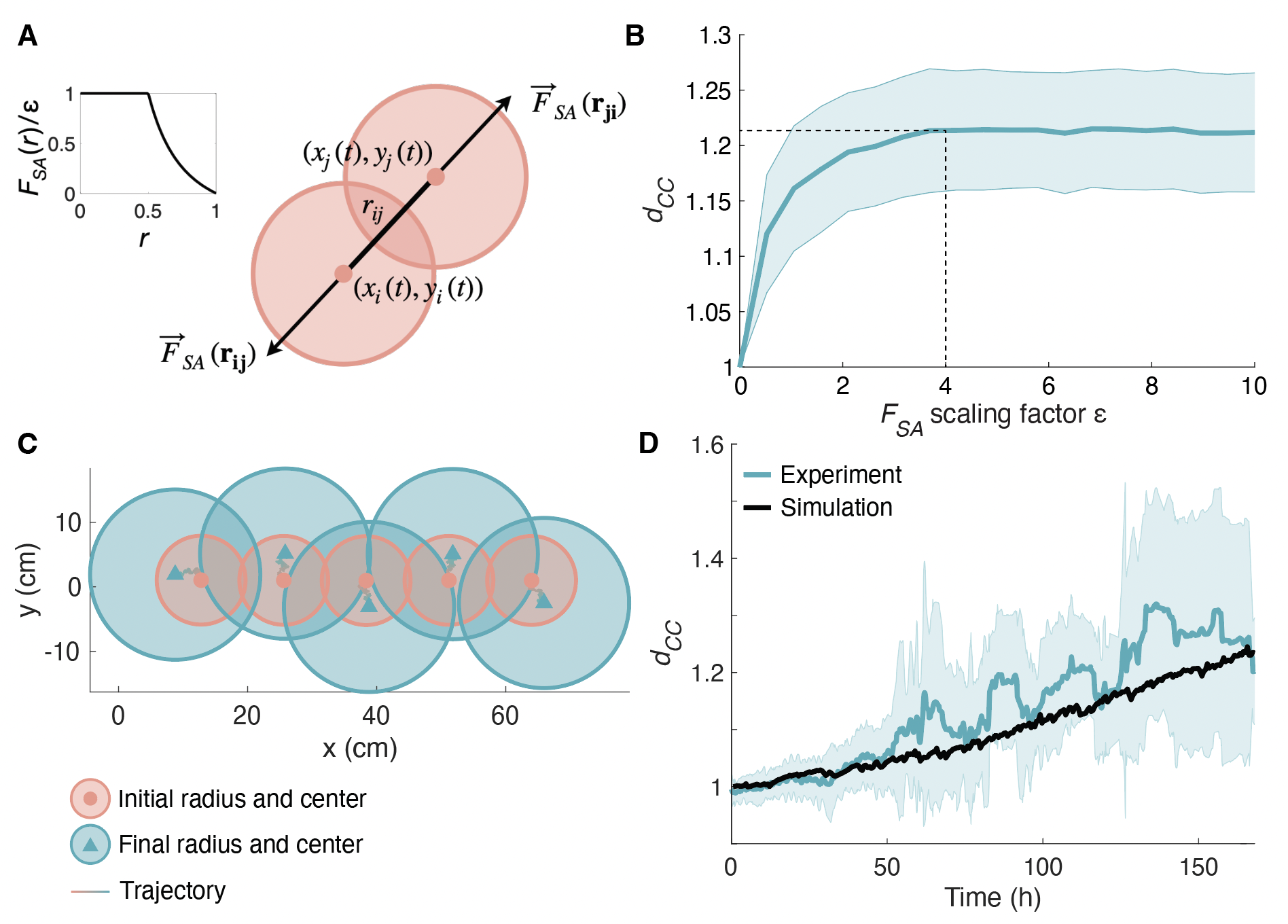}
    \caption{    \textbf{Plant-plant interaction simulations recover self-organization dynamics. (A)} In the model, two mutually-shading crowns separated by a distance $r_{ij}$ experience a repulsive force $F_{SA}$ with equal magnitude and opposite direction. The functional form of $F_{SA}$ as a function of crown separation $r$ is shown in the inset. 
    \textbf{(B)} The final center-center distance $d_{CC}$ varies as a function of the shade avoidance force scaling factor $\varepsilon$. Dashed lines indicate the value of $\varepsilon$, approximately 4, that corresponds to the experimental final $d_{CC}$ (panel E). 
    \textbf{(C)} Initial (pink circles) and final (blue circles) configurations of plant crowns over 7 simulated days, with centers marked by pink dots and blue triangles, respectively. 
    \textbf{(D)} The center-center distance ($d_{CC}$) between pairs of adjacent crowns increases over time. The average over 12 experiments is indicated by the blue line, with the shaded area representing 1 standard deviation. The black line shows $d_{CC}$ over the course of a simulation with shade avoidance scaling factor $\varepsilon = 4$ and step sizes sampled from the experimental distribution (Fig. \ref{fig:single_plant}F).} 
    \label{fig:sims1}
\end{figure}

We formulate a minimal model describing interacting plant crowns in the 2D plane which recapitulates the observed dynamics. 
We emulate growing plant crowns as circular disks whose area expands according to a measured growth rate (Fig.~\ref{fig:single_plant}A), and approximate the perturbations driven by circumnutations as a random walk, whose characteristics are determined from observed crown statistics (Fig.~\ref{fig:single_plant}C). 
When two crowns overlap, the shade avoidance response is represented by an effective repulsive force $F_{\text{SA}}$, scaled by a factor $\varepsilon$ (Fig.~\ref{fig:sims1}A).
All three approximations are informed by experimental data, as detailed in what follows.

\vspace{5 pt}
\noindent\textbf{Single plant dynamics.} 
We approximate a growing plant crown as a disk whose area $A(t)$ increases linearly in time with a growth rate $\vg = dA/dt$, graphically represented in Fig.~\ref{fig:single_plant}A. 
We evaluate crown size from single plant experiments over 10 days. Fig.~\ref{fig:single_plant}A shows an example of the segmentation of a plant crown, with a snapshot from the beginning of the experiment overlaid with one from the end. Fig.~\ref{fig:single_plant}B shows $A(t)$ the average area of the crowns as a function of time. A linear fit yields a crown growth rate of 
\begin{equation}\label{eq:growth}
\vg = \frac{dA}{dt} = 2.4 \pm 0.5 \;\frac{\text{cm}^2}{h}
\end{equation}

We approximate the movement of single plants, due to circumnutations, as a random walk (schematically shown in Fig.~\ref{fig:single_plant}C). The statistical characteristics of the random walk are evaluated from extracted trajectories of the crown centers from single plant experiments. An example of a trajectory spanning 10 days is shown in Fig.~\ref{fig:single_plant}C, where discontinuities are due to the untracked movement during nighttime (8 hours).  
In order to quantify the  circumnutation movements we first assess the sampling time step which captures the movement in a relevant way; choosing a time step which is too short over-samples the trajectory, resulting in correlated steps, and if it is too long it loses information.
We therefore calculate the step-step velocity auto-correlation $C_{vv}(t)$ (Eq.~\ref{eq:Cvv_methods}) as a function of different time steps (Fig. \ref{fig:single_plant}D), and choose the shortest time step at which the auto-correlation goes to zero, at $\Delta t =  0.6$ h. That is, steps $\Delta t =  0.6$ h apart are uncorrelated and can be modeled as a random walk. 
We further validate this by examining the mean squared displacement $\langle r^2(t) \rangle$ (MSD) of crown centers (Eq.~\ref{eq:MSD_methods}) averaged over 8 plants each recorded over a period of up to 10 days (Fig.~\ref{fig:single_plant}E). Gaps in the MSD represents the untracked nighttime movement. 
The MSD agrees with a linear fit, yielding 
\begin{equation}\label{eq:MSD}
\langle r^2(t) \rangle = 0.027t + 0.074 \;\text{cm}^2
\end{equation}
for $t$ expressed in hours, with coefficient of determination $R^2 = 0.84$. 

Next, we extract the distribution of step sizes, and find that it exhibits a remarkably broad range,
spanning almost 3 orders of magnitude. The distribution is very well approximated with a 2-component Gaussian mixture model (GMM), with $R^2 = 0.98$, as well as a truncated power law 
\begin{equation}\label{eq:step_dist}
\Px \propto \frac{1}{\Delta x^{\alpha}}
\end{equation}
with exponent $\alpha=1.38 \pm 0.05$ (Fig. \ref{fig:single_plant}F). The uncertainty of the exponent is determined by bootstrapping (Methods). Although the GMM fit is slightly better, we choose to utilize the power law form as it effectively captures the breadth of the distribution using a single parameter, which proves instrumental for the analysis conducted in the following section.

\noindent Lastly, we note that the auto-correlation exhibits small anti-correlations at time lags of $0.6-1.5$ h, reflecting the fact that crown movement is constrained by the length and flexibility of their stems, tethering them to the ground. 

This is consistent with the fact that while the MSD increases overall, reflecting the increasing range of motion due to the growing stems, the daily MSD averaged over single days exhibits a slight decrease of the slope after about 1 h (inset of Fig.~\ref{fig:single_plant}E). 
Indeed, simulating crown movement as a simple  uncorrelated random walk with steps taken from the extracted distribution, results in a faster MSD than the measured one. 
We therefore introduce a growing circular reflective boundary surrounding each crown in the model, with a radius $R_{B}(t)$ increasing linearly in time, whose rate is determined by tuning the simulated MSD to agree with the measured one:
\begin{equation}\label{eq:boundary}
R_{B}(t) = 0.023t + 0.074 \text{ cm}^2.
\end{equation}
Put together, we model fluctuations in crown position as a bounded uncorrelated random walk with steps taken from the extracted distribution, akin to a bounded truncated L\'{e}vy flight~\cite{Chechkin2008, mantegna1994}. Simulations recover the trend of the experimental MSD (Fig. \ref{fig:single_plant}E).

\begin{figure}[t!]
    \centering
    \includegraphics[width=0.7\linewidth]{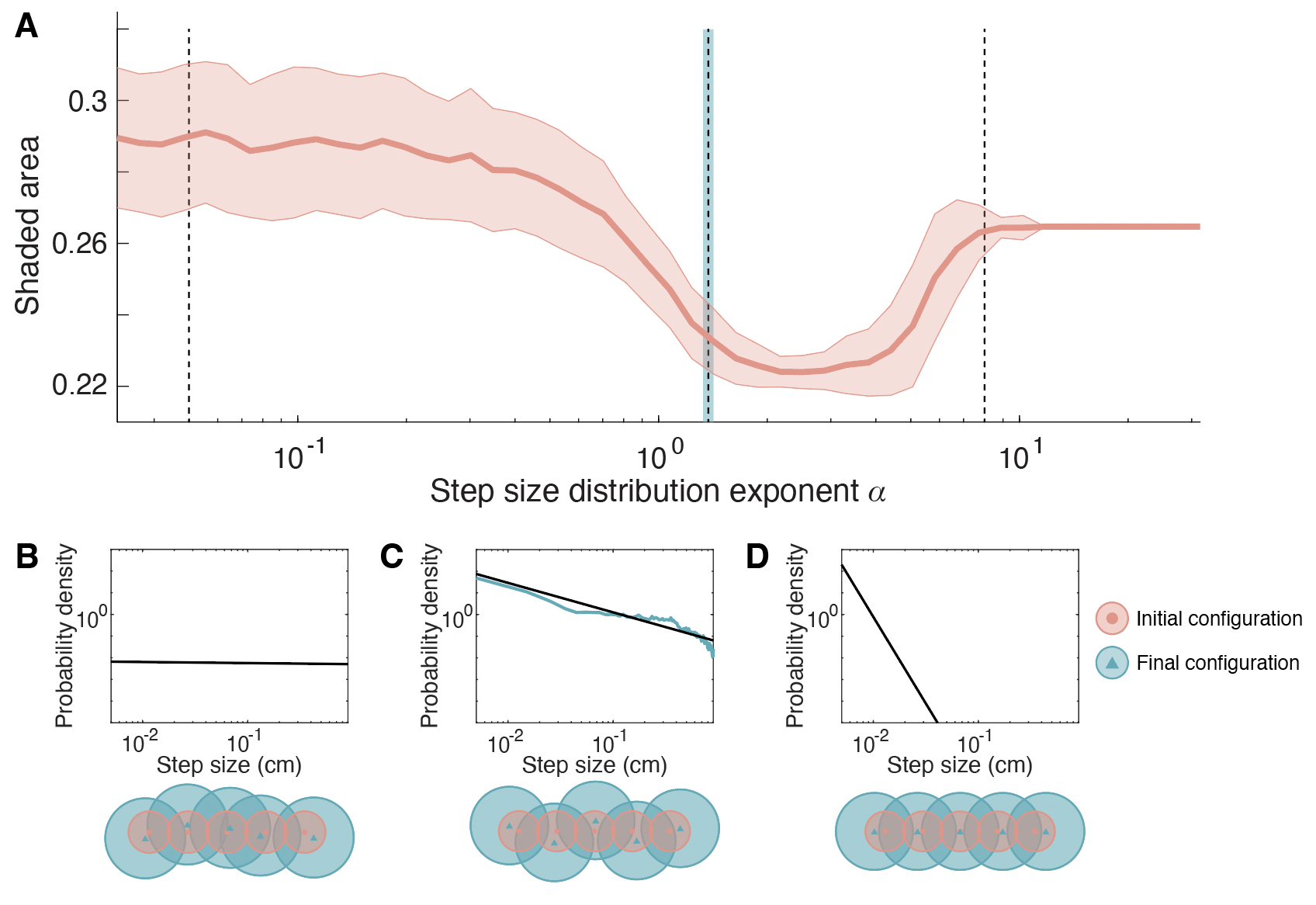}%}}
    \caption{\textbf{Simulations suggest circumnutations as source of functional noise}. 
    \textbf{(A)} Fraction of shaded area as a function of $\alpha$, the power-law exponent of the step size distribution in Eq.~\ref{eq:step_dist}. The center red line represents an average over 300 simulations, while the shaded area corresponds to a width of 2 standard deviations.
    Smaller values of $\alpha$ describe a wider distribution, representing more noise in the system. 
    The parameter sweep reveals three regimes: for high noise (small values of $alpha$), random movements dominate over repulsive interactions, producing disordered configurations that can result in greater shaded area. For very low noise (large values of $\alpha)$, plants remain close to their initial locations due to the lack of symmetry-breaking fluctuations that push plants off the horizontal axis. In between, there is a minimum at an optimal range of $\alpha$, where symmetry breaking results in the plants self-organizing into an alternately-deflecting ``zig-zag'' pattern. 
    \textbf{(B)-D} Examples of step-size distributions for $\alpha=8$ (low noise), $\alpha= 1.38$ (measured value), and $\alpha=0.05$ (high noise), respectively, with examples of simulated final plant configurations for each respective distribution.}
    \label{fig:sims2}
\end{figure}

\vspace{5 pt}
\noindent\textbf{Effective pairwise interaction.} 
We approximate the sunflower shade avoidance response, and other possible contributions such as mechanical interactions (see Video 1 in the Supplementary Material (SM)), as a pairwise interaction between disks. When two crowns overlap they experience an effective repulsive \textit{force} $F_{SA}$ (Fig.~\ref{fig:sims1}A) which, 
in the absence of a clear relation, we assume follows a simple inverse-square repulsion:
\begin{equation}\label{eq:Fsa}
\Fi^{ij}(\vec{r}_{ij})= -\varepsilon \left( \left(\frac{2R}{r_{ij}}\right)^2 - 1\right)\hat{r}^{ij}, \;\;\;\;\;  |\vec{r}_{ij}| < 2R.   
\end{equation}
Here $\varepsilon$ is a coefficient that scales the magnitude of the force and is determined from experiments, as detailed in the next section. 
$r_{ij} = |\vec{r}_{ij}| = |\vec{r}_{j} - \vec{r}_{i}|$ is the center-to-center distance between two crowns $i$ and $j$ at time $t$ (explicit time dependence omitted for clarity),  $\hat{r}_{ij} = \vec{r}_{ij}/{r}_{ij}$ is the unit vector, and  $\vec{F}_{SA}^{ji} = - \vec{F}_{SA}^{ij}$. With this form 
$\Fi^{ij}(\vec{r}_{ij}\rightarrow 2R) \rightarrow 0$, 
avoiding a jump discontinuity at $r_{ij} = 2R$. 
Furthermore, we fix $\vec{F}_{SA}^{ij} (\vec{r}_{ij}) = \vec{F}_{SA}^{ij}(R)$ for $r_{ij} < R$ to prevent extremely large forces in simulations. 
For a system of multiple mutually shading crowns, the force acting on crown $i$ is the sum of all pairwise forces with other crowns 
$\Fi^{i}(\vec{r}_{ij})= \sum_{j\neq i} {\Fi^{ij}(\vec{r}_{ij})}$. We later show that resulting dynamics are robust to different powers in Eq.~\ref{eq:Fsa} (Fig.~S2). 

\vspace{5 pt}
\noindent\textbf{Simulations recover measured self-organization dynamics.}
We use this model to simulate observed self-organization dynamics in experiments of 5 sunflower crowns (such as the one shown in Fig.~\ref{fig:plants_row}), initially aligned along the horizontal axis and placed $6.8$ cm apart, over the course of 7 days.   
To determine the scaling factor $\varepsilon$ of the shade avoidance force, we perform a parameter sweep, running simulations over a range of values of $\varepsilon$. For every simulation we calculate the final $d_{CC}$, and find that it generally increases for larger values of $\varepsilon$ (Fig.~\ref{fig:sims1}B),
thus capturing the general self-organization dynamics.
We identify that $\varepsilon \approx 4$ reproduces the experimental value $d_{CC} = 1.22$, as shown in Fig.~\ref{fig:sims1}C.
Put together with the crown growth rate and the single plant perturbation dynamics, despite the anticipated experimental spread our model captures the general trends %\cred{
of the evolution of both the MSD of single plants (Fig.~\ref{fig:single_plant}E), as well as $d_{CC}(t)$ in rows of multiple shading plants (Fig.~\ref{fig:sims1}C-D),  providing a quantitative description of the system's characteristic dynamics.

\subsection*{Analysis of effects of circumnutations on the self-organization process}

Having corroborated our minimal model, it now serves as a virtual laboratory, enabling us to interrogate the role of circumnutations, represented as noise, in the self-organization process. 
We quantify the amount of noise in the system with the width of the step size distribution. Assuming a general power-law form as in Eq.~\ref{eq:step_dist}, the width can be tuned with the exponent $\alpha$: large values of $\alpha$ correspond to narrow distributions and therefore mostly small fluctuations, while small values of $\alpha$ correspond to (truncated) heavy-tailed distributions allowing large fluctuations (illustrated in Fig.~\ref{fig:sims2}). 
We evaluate the performance of a system of self-organizing sunflowers in terms of the shaded area of the crowns after 7 days, relative to the total crown areas, termed SA (Eq.~\ref{eq:SA}): the lower the shaded area, the greater the energy the system is able to extract from photosynthesis, the better the performance.

We run simulations with all parameters set by experiments, as described before, and perform a parameter sweep over $\alpha$. 
Fig.~\ref{fig:sims2} displays the relative shaded area SA as a function of $\alpha$, and reveals three distinct regimes. 
For low perturbations with $\alpha \geq 8$, the shaded area remains constant. It then decreases to a minimum for intermediate perturbations with $8 < \alpha < 0.5$, and increases monotonically for larger fluctuations with $\alpha < 0.5$. 
While the distributions are truncated due to physiological limitations, it is interesting to note that non-truncated power-law distributions have a well-defined mean over 
$\Delta x\in [1,\infty )$ only if 
$\alpha >2$, and have a finite variance for 
 $\alpha>3$.

Examples of final configurations from simulations  corresponding to different regimes are shown in Fig.~\ref{fig:sims2}B-D: strong noise ($\alpha = 0.05$), moderate noise (the experimentally measured distribution, $\alpha = 1.38$), and weak noise ($\alpha = 8$). As expected, with weak noise plants remain close to their initial linear configuration, only deflecting along the row, but not away from it. This can be understood by considering that shading neighbors on opposite sides of a crown will convey repulsive forces of equal magnitude but opposite directions, resulting in no net movement of the crown away from the line. 

As noise increases, random fluctuations cause \textit{pioneer} plants to break the symmetry of the system by displacing off the horizontal axis, resulting in  alternating off-axis deflections in a zig-zag pattern that minimizes the shaded area.
This represents the ability of the system to explore a variety of states.  
As noise increases further, random movements dominate over the repulsive forces, potentially leading to sub-optimal disordered configurations with higher shaded area. We observe that the experimentally-determined noise, represented by the power law exponent of $\alpha = 1.38 \pm 0.05$, is within the range of minimal shaded area, demonstrating how sunflowers may leverage circumnutations as functional noise to self-organize optimally. 

Our simulations also reveal that the experimentally measured noise ensures that the process is robust to variations in the repulsive interaction,
which may correspond to, for example, variations in the shading due to leaf thickness, distance between leaves, or fluctuations in the environmental lighting (Fig.~S2).

\section*{Discussion}
 
While circumnutations are ubiquitous in plant systems, and generally associated with exploratory movements, a quantitative understanding of their role is elusive. Here we report, for the first time, their role in facilitating an optimal growth pattern for a crowded group of mutually shading plants.
We revisit the problem of self-organization observed for  sunflowers~\cite{Pereira2017}, mediated by shade response interactions, suggesting circumnutations as a source of functional noise. 
In order to test our hypothesis, we developed a parsimonious model of interacting growing disks, informed by experiments, successfully capturing the characteristic dynamics of single plants as well as multiple interacting plants. 
This framework provided an \textit{in silico} laboratory, enabling us to interrogate the role of circumnutations.

While traditionally circumnutation movements have  been investigated in terms of the geometry of the final trajectory, here we examined the characteristic statistics of their dynamics. We found that the movements can be described as a bounded random walk  characterized by a remarkably broad distribution of step sizes, or velocities, covering three orders of magnitude. 
This wide distribution may be related both to internal and external factors, such as ultradian and circadian rhythms, mechanical effects associated with self-weight, changes in turgor pressure due to watering, and changing light gradients.

We found that the observed breadth of the velocity distribution is beneficial, enabling the system to explore possible configurations in order to reach an optimum in terms of minimal shading, thus solving a explore-vs-exploit problem~\cite{Padilla2014, Wilson2021}. We therefore interpret circumnutations as \textit{functional} noise. 
Indeed, in motile animal systems such wide distributions of movement velocities are frequently identified with enhancement of behavioral processes, for example truncated power laws yielding L{\'e}vy flights, associated with animal search and foraging~\cite{VISWANATHAN2008133}, and broad shouldered distributions related to sensory salience~\cite{Biswas2023, Wilson2021}. 
We note that while we find here that circumnutations are beneficial, they also pose a cost to the plant, both due to the continuous change in leaf orientation (which by definition will not always be in the direction of light), as well as the mechanical cost of drooping sideways compared to growing straight. This cost-benefit trade-off needs to be addressed in future work.
To the best of our knowledge this is the first report of functional noise in plant movements, and provides a theoretical backdrop for investigating how plants negotiate their environment, employing computational processes such as task-oriented  processes, optimization, and active sensing.

\vspace{2 cm}

\section*{Methods}

\subsection*{Plant experiments}
We conducted single- and multiple-plant assays using sunflowers grown from seed (`EMEK 6' variety, Sha'ar Ha'amakim Seeds). The seeds were first cooled in a refrigerator (seed stratification) at $5$\textdegree C, peeled from their shell coats, and soaked in water for 24 hours. Each seed was then placed in a plastic test tube filled with wet Vermiculite and left to germinate in a growth chamber at 24{\textdegree}C, with a relative humidity of $72\%$ and a 12:12 h light:dark photoperiod. The light intensity in the chamber was 22.05 W/m$^2$.

Germination occurred after 4-7 days for each batch. Following germination, 3- to 7-cm-tall seedlings with two to four leaves that appeared healthy and well-separated were transplanted in $10$ cm $\times$ 10 cm-wide and 15 cm-tall black plastic pots containing garden soil. The plants were exposed to white LED light with intensity $41.92$ W/m$^2$ on a 16:8 h light:dark cycle, and the setup was enclosed with black fabric to eliminate sources of reflection. 
The ambient temperature was approximately 26{\textdegree}C during the light period and 28{\textdegree}C during the dark period, and the humidity ranged between 43-51\%. Each plant was watered with 100 mL of a 0.2\% 20-20-20 NPK fertilizer solution every 2 days. Plants were maintained in this setup for approximately one week.
 
Plants were then selected for single-plant or multiple-plant assays that took place in conditions similar to the growth conditions described above, with the exception of light intensity at $27.05$ W/m$^2$. In single-plant assays (9 experiments in total), plants were maintained in the enclosed experimental setup for 7-10 consecutive days. Multiple-plant assays consisted of five plants in individual pots closely arranged side-by-side in a horizontal row, again for 7-10 consecutive days (Fig.~\ref{fig:plants_row}A-B). Out of the total 12 multi-plant experiments 6 were on a continuous light regimen, however the characteristic statistics are similar to 16:8 light:dark regimen (Fig.~S1). 

\subsection*{Image acquisition and tracking}

We record plants during a 16-hour-long ``light'' period for up to 10 consecutive days. Each plant assay was recorded from a top-down view with a Logitech C270 HD webcam. Images were acquired every 5 minutes during the light period using a Raspberry Pi Model 4 single-board computer. We obtain images from 8 single-plant assays and 13 multiple-plant assays.

We perform image segmentation on videos of plants grown in individual setups using the colorThresholder function in MATLAB (MathWorks Inc., Natick, MA), from which we determine the area of the plant crown from the top-down view in single-plant assays. We segment the plant crown for every image, thus recording the crown area and the crown's center position as a function of time.

For multiple-plant assays, plant crowns can overlap from the top-down views, complicating image segmentation. To track the movement of the crowns in both single- and multiple-plant experiments, we manually annotate the center point of each crown in the first frame of the video and track its position using the DLTdv tracking software \cite{DLTdv}.

\subsection*{Plant dynamics}

\textbf{Center-to-center distances between plants.} 
We define the position of crown $i$ at time $t$ as $\vec{r}_i(t)$, and the initial position as $\vec{r}^0_i = \vec{r}_i(0)$. We define the average center-to-center distance between adjacent plants $d_{\text{CC}}(t)$, normalized by the initial center-center distance at $t=0$, as
\begin{equation}\label{eq:c2c}
    d_{\text{CC}}(t) = \sum_{i=0}{|\vec{r}_i(t) - \vec{r}_{i+1}(t)|} / \sum_{i=0}{|\vec{r}^0_i - \vec{r}^0_{i+1}|}
\end{equation} 
This value serves as a measure of how closely ($d_{\text{CC}}< 1$) or sparsely ($d_{\text{CC}}> 1$) the plant crowns are distributed, where $d_{\text{CC}}(0) = 1$ by definition.

\textbf{Velocity autocorrelation.} 
To obtain a sampling rate of plant movement that captures random walk dynamics, we first quantify the autocorrelation of velocities, $C_{vv}(t)$. If the instantaneous velocity $\vec{v}(t)$ of the plant at time $t$ is given by $\vec{v}(t) = (\vec{r}(t) - \vec{r}(t-dt))/dt$, where $dt$ is the length of time between subsequent frames in the recording (5 minutes), then the velocity autocorrelation as a function of time lag $\Delta t$, $C_{vv}(
\Delta t)$, where $\Delta t = ndt$ is an integer $n$ multiple of the time between frames $dt$, is given by

\begin{equation}\label{eq:Cvv_methods}
    C_{vv}(\Delta t) = \frac{1}{(N-n)}\sum_{t = 0}^{Ndt}  \vec{v}(t) \cdot \vec{v}(t+\Delta t),
\end{equation}
where $N$ is the total number of frames in the video.
We observe that $C_{vv}(\Delta t)$ is initially positive for $\Delta t < 0.6 $h, before decreasing to 0 at $\Delta t \approx 0.6$ h. Hence we choose a time step in our model that corresponds to 0.6 h, such that steps taken 0.6 h apart are uncorrelated and can be represented by random walk.

\textbf{Step size distribution.} 
To determine the step size distribution, we compute the distribution of all steps separated by $\Delta t = 0.6$h, i.e. the distribution of $\Delta x(t) = |\vec{x}(t) - \vec{x}(t - 0.6\text{h})|$ for all $t \in [0, T]$ where $T$ is the length of the video. We perform a least-squares fit of this distribution to a power law, $P(\Delta x) \sim \Delta x^{-\alpha}$, and find a best-fit value of $\alpha = -1.4$. To obtain an error estimate of $\alpha$, we perform bootstrapping by sampling from the step size distribution with replacement, fitting the resampled distribution to a power law, and repeating the resampling and fitting process for $10^4$ iterations. We then compute the standard deviation of $\alpha$ over these $10^4$ bootstrapped iterations to obtain a value of 0.05.

\textbf{Mean squared displacement.} 
To characterize the trajectories of plants, we compute the mean squared displacement (MSD) across all single plant recordings as 
\begin{equation}\label{eq:MSD_methods}
    r^2(t) = \frac{1}{N_p} \sum_{i=1}^{N_p} |\vec{r}_i(t)-\vec{r}_i(0)|^2
\end{equation}
where $\vec{r}_i(t)$ is the position of plant $i$ at time $t$, and $N_p$ is the total number of plants, in this case $N_p = 8$.
There are gaps in the MSD during the nighttime periods when the plants were not recorded.
We perform a least-squares linear fit of the MSD, obtaining the best-fit expression $\text{MSD}(t) = (0.027 \text{ cm}^2\text{/h})t + 0.074 \text{ cm}^2$, for $t$ expressed in hours, with coefficient of determination $R^2 = 0.84$.

\subsection*{Minimal model of shade avoidance response}
We formulate a model of crown deflections in the 2-D plane by modeling each plant as a circular crown with an area that grows linearly in time. When the separation between a pair of crowns is less than the sum of their radii, each crown experiences a repulsive ``shade avoidance'' force $F_{SA}$ of equal magnitude (Fig.~\ref{fig:sims1}A).

For a system consisting of two crowns $i$ and $j$ each with radius $R$ and separated by the vector $\vec{r}^{\;ij} = \vec{r}^{\;j} - \vec{r}^{\;i}$, the shade avoidance force $\vec{F}_{SA}^{\;i}$ (Fig.~\ref{fig:sims1}A) acting on crown $i$ is given by
\begin{equation}\label{eq:Fsa_methods}
\vec{F}_{SA}^{\;i} (\vec{r}^{\;ij})= 
\begin{cases} 
-\varepsilon \left( \frac{(2R)^2}{|\vec{r}^{\;ij}|^2} - 1\right)\hat{r}^{\,ij} & |\vec{r}^{\;ij}| < 2R\\
     
      0 & |\vec{r}^{\;ij}| \geq 2R
   \end{cases},
\end{equation}
where $\varepsilon$ is a coefficient that scales the magnitude of the force, $\hat{r}^{\,ij}$ is the unit vector in the direction of $\vec{r}^{\;ij}$, and the $-1$ term is introduced to shift the value of the force such that it approaches 0 when $|\vec{r}^{\;ij}| \rightarrow (2R)^-$ and avoids the jump discontinuity that would otherwise occur at $|\vec{r}^{\;ij}| = 2R$. Furthermore, we fix $\vec{F}_{SA}^{\;i} (\vec{r}^{\;ij}) =\vec{F}_{SA}^{\;i} (0.5\vec{r}^{\;ij}) $ for $|\vec{r}^{\;ij}| < R $ to prevent extremely large forces from occurring in the simulation.

The shade avoidance force $\vec{F}_{SA}^{\;j}$ acting on crown $j$ is of equal magnitude and points in the opposite direction along the unit vector $\hat{r}^{\,ji}$.
For a system of multiple mutually shading crowns, the force $\vec{F}_{SA}^{\;i}$ acting on crown $i$ is the sum of all pairwise forces between crown $i$ and all other crowns in the system.

Then, at each time step of the simulation, the change in position of crown $i$ is given by the overdamped equation of motion
\begin{equation}\label{eq:motion}
    \vec{x}^{\;i}(t + \Delta t) = \vec{x}^{\;i}(t) + (\vec{F}_{sp}^{\;i} + \vec{F}_{SA}^{\;i}) \Delta t + \vec{\xi},
\end{equation}

where $\vec{\xi}$ is a vector representing a random step with direction sampled from the uniform distribution $[0, 2\pi)$ and magnitude sampled from the step size distribution extracted from single plant experiments (Fig.~\ref{fig:single_plant}D).

Because the center of the crown is rooted to the ground via a stem of finite length and stiffness, the crown's movement is constrained in the x-y plane by the amount by which the stem is able to deflect. To quantify this constraining region, we compute the mean squared displacement (MSD) of the crown over the course of entire single-plant experiments (Fig.~\ref{fig:single_plant}C). 
The MSD then represents the increasing radius of a circle, centered on the crown, that bounds the movement of the crown as the stem grows, allowing for greater deflection over time. The bounding circle acts as as a reflecting boundary condition: if the change in position of the crown places the crown center outside of this bounding circle, the excess movement outside the bounding circle is reflected back toward the interior of the circle.

The radius of the crown increases according to the relation
\begin{equation}
    R(t) = R_0 + \sqrt\frac{At}{\pi}
\end{equation}
where $R_0$ is the radius at the start of the simulation and $A$ is the growth rate of the crown. For simplicity, in our simulations, we set the radius to be equal across each crown in the system.

 We simulate multiple-plant systems with our model using experimentally-determined parameters. 5 plants are initially separated by a distance of 12.8 cm, the average value of the initial center-center distance $d_{CC}$ without normalization. Each crown has an initial radius of 6.8 cm, which is estimated by segmenting the plant area in the first frames of multiple-plant assays, and taking this segmented area to represent five circular crowns. The area of each crown increases linearly at a rate of 57 cm$^2$ (Fig.~\ref{fig:single_plant}B) per day. Each timestep of the simulation represents 0.6 h, and the magnitudes of their random movements are drawn from the experimental step size distribution (Fig.~\ref{fig:single_plant}E).

 The system performance is evaluated by determining the fractional shaded area across all plants, given by
 \begin{equation}\label{eq:SA}
     SA = \frac{1}{N_p \pi R^2}\sum_{i=1}^{N_p} \sum_{j=i+1}^{N_p} \left(2R^2 \cos^{-1} \left(\frac{r_{ij}}{2R}\right) - \frac{1}{2} r_{ij} \sqrt{4R^2-r_{ij}^2}\right)
 \end{equation}
 where $N_p$ is the total number of plants and $r_{ij}$ is the distance between the centers of plants $i$ and $j$.
\section*{Conflict of Interest}
The authors declare that they have no competing financial interests.

\section*{Funding}
O.P. and Y.M. acknowledge support from the Human Frontiers Science Program (HFSP), Young Investigator Grant RGY0078/2019. O.P. acknowledges support from the Army Research Office Grant 78234EG. Y.M. acknowledges support from the Israel Science Foundation Research Grant (ISF) no. 2307/22.

\end{document}